\documentclass[conference, letter]{IEEEtran}
\usepackage{cite}

\ifCLASSINFOpdf
   \usepackage[pdftex]{graphicx}
\else
  \usepackage[dvips]{graphicx}
  \DeclareGraphicsExtensions{.eps}
\fi
\usepackage[cmex10]{amsmath}
\usepackage{tikz}
\usetikzlibrary{matrix,shapes,arrows,positioning,chains}

\usepackage{amssymb}
\usepackage{mathtools}

\usepackage{color}
\usepackage{algorithm,algpseudocode}
\usepackage{caption}

\usepackage{varwidth}
\usepackage{fancybox,framed}
\usepackage{flushend}
\usepackage{caption}
\usepackage{subcaption}
\usepackage{fixltx2e}

\usepackage{stfloats}
\usepackage[left=1.62cm,right=1.62cm,top=1.9cm]{geometry}
\makeatletter

\newcommand{\Rmnum}[1]{\expandafter\@slowromancap\romannumeral #1@}
\makeatother

\newcounter{lemmano}
\newcounter{propono}

\begin{document}
%
\title{A canonical correlation-based framework for performance analysis of radio access networks
}


\author{
\IEEEauthorblockN{\textsuperscript{} Furqan Ahmed} 
\IEEEauthorblockA{
\textit{Elisa Corporation}\\
Helsinki, Finland \\
furqan.ahmed@elisa.fi}

\and

\IEEEauthorblockN{\textsuperscript{} Muhammad Zeeshan Asghar}
\IEEEauthorblockA{
\textit{Aalto University}\\
Espoo, Finland \\
muhammad.asghar@aalto.fi}

\and
\IEEEauthorblockN{\textsuperscript{} Jyri~H\"am\"al\"ainen}
\IEEEauthorblockA{
\textit{Aalto University}\\
Espoo, Finland \\
jyri.hamalainen@aalto.fi}
}


\maketitle
\begin{abstract}
Data driven optimization and machine learning based performance diagnostics of radio access networks entails significant challenges arising not only from the nature of underlying data sources but also due to complex spatio-temporal relationships and interdependencies between cells due to user mobility and varying traffic patterns.
We discuss how to study these configuration and performance management data sets and identify relationships between cells in terms of key performance indicators 
using multivariate analysis. To this end, we leverage a novel framework based on canonical correlation analysis (CCA), which is a highly effective method for not only dimensionality reduction but also for analyzing relationships across different sets of multivariate data. As a case study, we discuss energy saving use-case based on cell shutdown in commercial cellular networks, where we apply CCA to analyze the impact of capacity cell shutdown on the KPIs of coverage cell in the same sector. Data from LTE Network is used to analyzed example case. We conclude that CCA is a viable approach for identifying key relationships not only between network planning and configuration data, but also dynamic performance data, paving the way for endeavours such as dimensionality reduction, performance analysis, and root cause analysis for performance diagnostics.

\end{abstract}

%
\IEEEpeerreviewmaketitle

\section{Introduction}
%
%
%
%
\IEEEPARstart{R}adio access networks are becoming increasingly complex due to a multitude of factors including network densification, multiple radio access technologies, new frequency bands and spectrum sharing scenarios, advanced multi-antenna techniques, and stringent requirements for energy efficient operation based on heterogeneous user demands and skewed traffic patterns. This calls for a higher degree of network intelligence capable of analyzing massive amounts of data and computing correct values of a large number of parameters on-the-fly.
Computing exact solutions to underlying mathematical optimization problems is often prohibitively complex due to high dimensionality, and coupling constraints between problem variables. 
However, due to recent advances in cloud computing technologies, open source tools, and particularly machine learning and artificial intelligence algorithms and platforms, soft computing algorithms based on heuristics and meta-heuristics and data driven optimization approaches in general are garnering notable interest\cite{qc_ahmed}. The performance of such solutions is dependent on the computational and data processing capabilities of the underlying platform.
A key challenge, particularly relevant to commercial radio access networks, is the near real time processing and analysis of massive amounts of network data often characterized by challenges related to data volume, variety, velocity, and veracity. Apart from network optimization and automation use-cases, the use of data driven intelligent approaches are also applicable to performance monitoring and diagnostics, network anomaly detection and root cause analysis.

Existing approaches are motivated by the fact that analyzing massive amounts of data generated in mobile networks necessitates the use of data reduction techniques. A commonly used approach is to monitor performance metrics aggregated over cells and/or time. Degradation of KPIs triggers ticket creation, followed by detailed root cause analysis by domain experts. Although, aggregation reduces data, the process is highly time consuming and error prone. Moreover, aggregation often results in loss of local features, thereby impacting the quality of decisions. Recently, a number of approaches have been proposed for reducing the amount of data, thereby extracting only what impacts the performance. These include feature selection and feature extraction approaches \cite{shf_barco,de2020feature}, and other unsupervised techniques such as self-organizing maps and clustering\cite{ulbarco}. In \cite{shf_barco}, authors propose feature selection and feature extraction based dimensionality reduction framework for root cause analysis, whereas \cite{de2020feature} shows the practical application of such techniques as an intermediate stage between performance monitoring and network management functions. An automatic root cause analysis system based on self-organizing maps, clustering, and labelling is discussed in\cite{ulbarco}. Moreover, big data enabled network management use-cases have recently been discussed in literature. Notable examples include mobility robustness optimization\cite{joseph_mro}, anomaly detection\cite{moysen_bigdata2020}, and traffic forecasting\cite{xu}, to name a few. Canonical correlation analysis is a tool that can be used to study complex relationships between variables in multiple data sets. In particular, it can simultaneously evaluate multiple related data sets without any assumptions on directionality. It is a step beyond Pearson correlation (one to one) and many regression (many to one) analysis. In the context of big data analytics, the CCA has recently been garnering attention in neuroscience\cite{wang_neuro_cca}, data privacy\cite{wang_xmedia_cca}, and machine learning applications in general\cite{patil_dl_cca}. However, this approach is not well studied in the area of wireless networks, performance monitoring and RAN analytics in particular. To the best of our knowledge, apart from \cite{celledge_cca}, this approach has not been studied for RAN analysis.

In this paper, we discuss how to analyze RAN configuration and performance data sets and extract important parameters and counters using CCA. Having multiple sets of data on the same network element paves the way for CCA between variables. We discuss energy saving use-case based on cell shutdown feature as an example use-case, and study the impact of capacity cell shutdown on coverage cell performance data, identifying relevant performance counters to monitor once parameters are optimized.
The rest of the paper is organized as follows: Section~\Rmnum{2}
introduces CCA methods, and Section~\Rmnum{3} discusses how these can be applied to RAN data sets followed by a discussion on the preliminaries of quantum computing and proposes an architecture for quantum computing enabled mobile network automation platform. In Section~\Rmnum{4}, the proposed framework is applied to energy saving use-case to analyze the impact on coverage layer performance data. Numerical results from a commercial LTE network are discussed. Finally, conclusions are given in Section~\Rmnum{5}.

\section{Canonical Correlation Analysis}
\subsection{Introduction}
Canonical correlation can be used to explore relationship between two sets of high dimension variables (e.g. multivariate sets of variables) that are generated by the same system or process. The motivation for this approach is the inherent difficulty in analyzing relationships between such data sets, especially when the number of variables is high.
For instance, if there are $p$ variables in one set and $q$ in another set, total number of correlations to analyze becomes $pq$. Canonical correlation analysis reduces the dimensionality of the problem without losing any critical information, thereby providing relevant and summarized statistics that are easy to interpret. 
\subsection{Model}
Consider two sets $\mathcal{X} = \{X_1, X_2, \dots X_p\}$ and $\mathcal{Y} = \{Y_1, Y_2, \dots Y_q\}$, comprising of $p$ and $q$ variables respectively. For each set, we can reduce the variables by using aggregation based on weighted linear combinations. The aggregation can be considered as a function of the set of variables. This function is usually a linear function, and resulting variable is known as canonical variable as it represents the underlying set of variables. The idea is to find relationship between the aggregated variables such that the correlation between sets is maximized. Let $\mathbf{A}=[\mathbf{a}_1, \mathbf{a}_2, \dots \mathbf{a}_p]$ be the coefficients for variables in set $\mathcal{X}$, where $\mathbf{a}_1 = [{a}_{11}, {a}_{21}, \dots {a}_{p1}]^T$. Then, elements of set $\mathcal{U} = {U_1, U_2, \dots ,U_p}$ comprising of
canonical variables for set $\mathcal{X}$ be expressed as:

$$
\mathcal{U} \coloneqq \left\{
\begin{array}{cc}
  U_1 &= a_{11}X_{1} + a_{12}X_{2} + \dots + a_{1p} X_{p} \\
  U_2 &= a_{21}X_{1} + a_{22}X_{2} + \dots + a_{2p} X_{p} \\
      & \vdots \\
  U_p &= a_{p1}X_{1} + a_{p2}X_{2} + \dots + a_{pp} X_{p} 
\end{array}
\right\}
$$

We assume $p \le q$, and cardinality of set $\mathcal{U}$ and $\mathcal{V}$ is $\min {(p,q)}$. Next, for set $\mathcal{Y}$, we define set $\mathcal{V} = {V_1, V_2, \dots ,V_p}$, where $\mathbf{B}=[\mathbf{b}_1, \mathbf{b}_2, \dots \mathbf{b}_p]$, and $\mathbf{b}_1 = [{b}_{11}, {b}_{21}, \dots {b}_{p1}]^T$. The elements of $\mathcal{V}$ are:

$$
\mathcal{V} \coloneqq \left\{
\begin{array}{cc} 
  V_1 &= b_{11}Y_{1} + b_{12}Y_{2} + \dots + b_{1q} Y_{q} \\
  V_2 &= b_{21}Y_{1} + b_{22}Y_{2} + \dots + b_{2q} Y_{q} \\
  & \vdots \\
  V_p &= b_{p1}Y_{1} + b_{p2}Y_{2} + \dots + b_{pq} Y_{q} \\
\end{array}
\right\}
$$

Each element of $\mathcal{U}$ is paired with an element of $\mathcal{V}$. First canonical variate pair is $(U_1, V_1)$. Likewise, $(U_i, V_i)$ denotes $i^{\text {th}}$ pair. Each set of weights ${a}_{i1}, {a}_{i2}, \dots {a}_{ip}$ and ${b}_{i1}, {b}_{i2}, \dots {b}_{iq}$, gives $(U_i, V_i)$ pair. Correlation between $U_i$ and $V_i$ is called $i^{\text {th}}$ canonical correlation and can be calculated as
\begin{equation*}
  \rho = \frac{\text{cov}(U_i,V_i)}{\sqrt{\text{var}(U_i)\text{var}(V_i)}}
\end{equation*}

The aim is to find coefficients that maximize $\rho$ under certain constraints. For instance, constraints for maximizing the first pair is $\text{var}(U_1)=\text{var}(V_1)=1$. For each pair, the aim is to find coefficients that maximize the correlation between the members of pair. It is important to note that first pair $(U_1, V_1)$ is the most important one, as it corresponds to the maximum correlation. For the second pair, there are additional constraints that $(U_1, U_2)$, $(V_1, V_2)$, $(U_1, V_2)$, and $(U_2, V_1)$  are uncorrelated. For the sake of brevity, details on subsequent pairs are omitted. These can be found in any standard text on the subject~\cite{thompson2000canonical}.

\section{CCA for Mobile Network Data sets}
Different types of data sets are available in commercial mobile networks, which are useful for not only network optimization use-cases, but also for fault diagnostics and root cause analysis. These include configuration management (CM), performance management (PM), inventory management (IM), and fault management (FM). For standard SON use-cases, data is fetched from the network on a daily basis, and is stored in the environment for daily KPI monitoring as well as tuning network parameters.

\subsection{CM Data}
The CM data consists of all the configuration parameters such as identifiers (PCI, scrambling codes etc), random access parameters, and mobility parameters, frequency bands and channels, existing neighbor relations between cells etc. Its analysis provides information regarding current network parameters and configuration. Some of these parameters change frequently as a result of network optimization activities.

\subsection{PM Data}
PM data consists of a number of different performance counters collected by the network. Compared to CM data, it changes very frequently collecting related data from the network at regular intervals with a very high granularity (e.g. 15 minutes to one hour). Examples include cell level counters that collect statistics for the whole cell, and cell pair level counters that collect statistics from the source cell to each neighboring cell. Cell pair counters require much more storage and computing resources. In addition, there are board counters that collect different board level measurements.

\subsection{IM Data} 
IM data consists of information related to physical resources (cabinets, subracks, slots, boards, antennas, GPS etc.) and logical resources (e.g. versions) of managed network elements, which helps operators to manage these resources in an efficient manner. Examples include cell site and antenna related information such as geographical location, antenna type, and bearings. This data is usually static in that it is usually not changed during operation. It is independent from the operating status of equipment and CM data. Moreover, it includes physical relationships between resources and unlike CM the parameter values cannot be changed.

\subsection{CCA Analysis}
\begin{figure}[t]
\begin{center}
\includegraphics[trim={0cm 0cm 0cm 0cm},scale=0.35,clip]{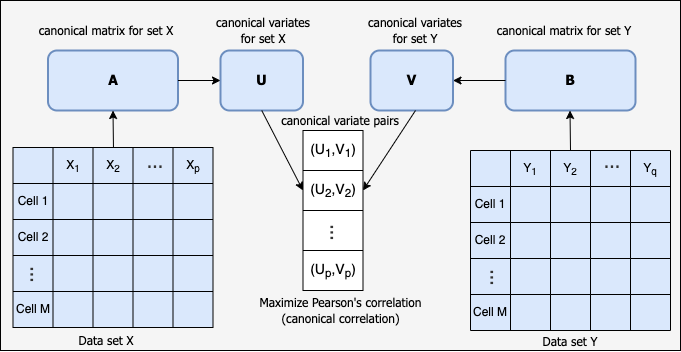}
\end{center}
\caption{CCA for finding relationships between different cell level variables in the full network.}
\vspace{0mm}
\label{fig:framework1}
\end{figure}

\begin{figure}[t]
\begin{center}
\includegraphics[trim={0cm 0cm 0cm 0cm},scale=0.35,clip]{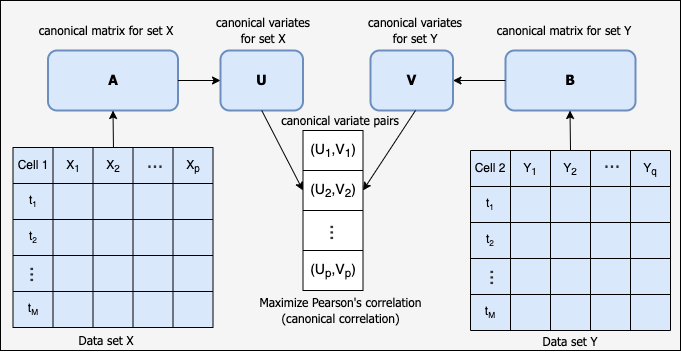}
\end{center}
\caption{CCA for finding relationships between different cell level variables belonging to different cells in the same sector.}
\vspace{0mm}
\label{fig:framework2}
\end{figure}

In order to apply CCA for RAN data analysis, we need two related sets of structured data comprising of multiple variables. These variables could belong to any of the categories discussed above. In practice, aforementioned three data sources are consolidated into a unified model, which is then used for implementing network optimization algorithms. Having such model is particularly important for complex algorithms where decisions are based not only on current CM parameters but also IM data and historical PM data.
The concept of CCA is illustrated in Fig.~\ref{fig:framework1}, where two data sets X and Y are shown. These data sets belong to the same set of cells given in first columns. Thus, cell ID columns are exactly the same. However, rest of the columns possibly representing (CM/PM/IM) variables are different. The aim is to analyze relationships between variables across the sets. These variables could be CM parameters or PM counters aggregated over certain time period depending on the use-case.
Figure~\ref{fig:framework2} illustrates another example, consisting of time-series data. In this case, two sets belong to two different cells, however these two cells are related in that they are in the same sector with overlapping coverage. Therefore, they are coupled together and highly likely to impact each other. Also, in this case too the two sets have same time stamps so measurements are aligned, which will give a clear picture of relationships over this time horizon. 
\section{CCA of Cell Shutdown Based Energy Saving}
We consider RAN energy saving use-case to discuss CCA for performance analysis. The motivation of using this example is that it consists of multiple CM parameters and PM counters that are closely related and therefore impact each other. Understanding the impact of shutdown on different KPIs related to cell performance and user experience is an important problem.

\subsection{Energy Savings using Capacity Cell Shutdown}
A key approach for saving energy in inter-frequency co-coverage networks during off-peak hours when network load is light is to handover users to inter-frequency co-coverage neighbour cell and turn-off the capacity cell. Such features are provided by vendors and can be switched on by configuring relevant parameters. The decision to turn-off capacity cell is based on number of users and physical resource block (PRB) usage in coverage and capacity cells. Shutdown timings are determined based on the PRB availability in coverage cell and PRB usage in the capacity cells, in both uplink and downlink. CM parameters such as thresholds on PRB usage and number of users are used to adjust the shutdown behaviour. A cell enters the shutdown state when the sum of downlink PRB usage of the capacity cell and coverage cell is less than the downlink PRB threshold. Similar condition is used for the uplink. In addition, the number of users in the capacity cell is checked. The shutdown happens only if this number is less than number of users threshold. A number of PM data counters can be used for analyzing the performance of energy saving schemes. These include cell shutdown statistics and other counters for measuring cell performance and user experience.

\subsection{CCA Model and KPIs}
\begin{figure} 
    \centering
  \subfloat[Hourly downlink PRB usage.\label{1a}]{%
       \includegraphics[trim={0.0cm 0cm 0cm, 0.0cm},scale=0.3,clip]{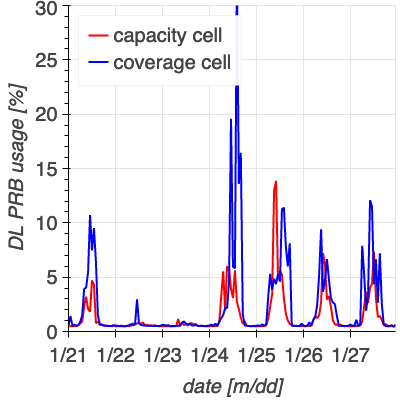}}
    \hfill
  \subfloat[Hourly uplink PRB usage.\label{1b}]{%
        \includegraphics[trim={0.0cm 0cm 0cm 0cm},scale=0.3,clip]{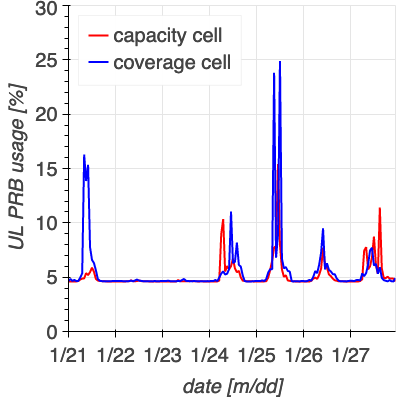}}
  \caption{Downlink and uplink PRB usages for one week.}
  \label{fig:prb_usage} 
\end{figure}

\begin{figure} 
    \centering
  \subfloat[Hourly unavailability data.\label{1a}]{%
       \includegraphics[trim={0.0cm 0cm 0cm, 0.0cm},scale=0.3,clip]{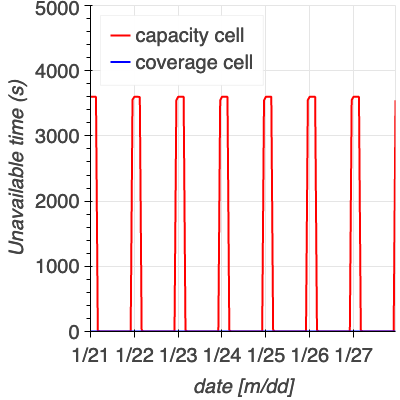}}
    \hfill
  \subfloat[Hourly max. DL TX power.\label{1b}]{%
        \includegraphics[trim={0.0cm 0cm 0cm 0cm},scale=0.3,clip]{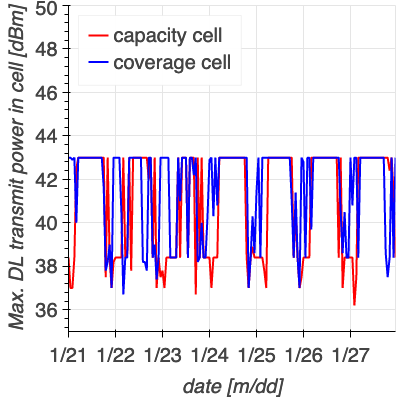}}
  \caption{Cell unavailability and transmit powers.}
  \label{fig:unavail_power} 
\end{figure}

\begin{figure} 
    \centering
  \subfloat[Hourly downlink throughput.\label{1a}]{%
       \includegraphics[trim={0.0cm 0cm 0cm, 0.0cm},scale=0.3,clip]{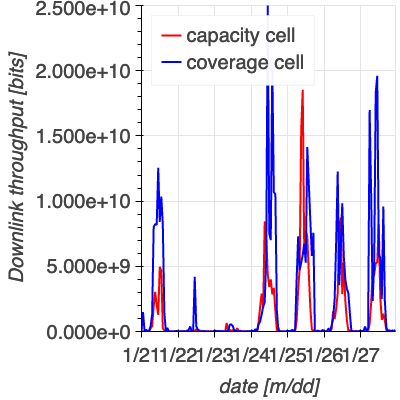}}
    \hfill
  \subfloat[Hourly avg. number of users.\label{1b}]{%
        \includegraphics[trim={0.0cm 0cm 0cm 0cm},scale=0.3,clip]{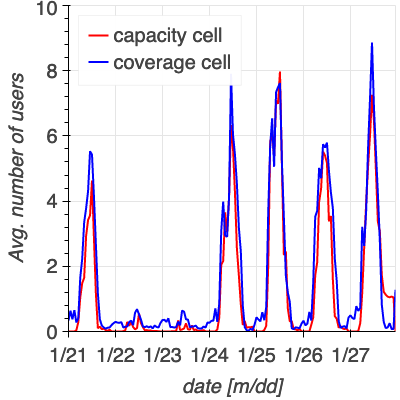}}
  \caption{Throughput and number of users.}
  \label{fig:tp_users} 
\end{figure}

Let us consider sector $S$ in a cell site comprising of multiple sectors. Sector $S$ consists of $N$ cells $c_1, c_2, \dots , c_N$ operating on different frequencies. As the coverage area of cells overlap, we assume that certain parameter changes in one cell are likely to impact performance data of other cells in the same sector in addition to the changed cell's own KPIs. Each cell $c_n$ has a KPI data set comprising of $K$ time series, each with $M$ elements. Without the loss of generality, let us consider a pair of cells $c_x$ and $c_y$ where $c_x$ is the cell that changes its configuration parameter and $\{c_y\}_{y\neq{x}}$ is another cell in the same sector which may be impacted by the actions of cell $c_x$. We are interested in analyzing the relationship between KPIs of $c_x$ and $c_y$. This will lead to insights on how to mitigate the impact of $c_x$ actions on other same sector cells.
To this end, we consider $p$ KPIs for $c_x$ and $q$ for $c_y$. Resulting data sets $\mathbf{X}_{M \times p }$ and $\mathbf{Y}_{M \times q }$, corresponding to $c_x$ and $c_y$, respectively. 
Note that $\min(p,q)$ linear combinations and canonical pairs are possible.
The main idea behind this approach is find coefficients $\mathbf{a}$ and $\mathbf{b}$ such that correlation between linear combination of $p$ column elements of $\mathbf{X}$ with linear combination of $q$ column elements of $\mathbf{Y}$ is maximized. Then, canonical correlation $\rho_{xy}$ between $\mathbf{X}_{M \times p }$ and $\mathbf{Y}_{M \times q }$ can be expressed as:

\begin{equation*}
  \rho_{xy} = \frac{\mathbf{X}\mathbf{u}\cdot\mathbf{Y} \mathbf{v}}{{\|\mathbf{X}\mathbf{u}\|}_{2} \cdot {\|\mathbf{Y}\mathbf{v}\|}_{2}}
\end{equation*}
We consider some standard KPIs relevant to energy savings use-case such as downlink and uplink PRB usages, shutdown or unavailable times, maximum downlink transmit power, downlink throughput, and average number of users. The KPI data, taken from a commercial LTE network, has hourly granularity and duration of one week. Fig. \ref{fig:prb_usage} shows PRB usages in both downlink and uplink for capacity and coverage cells. It can be seen that usage drops during off-peak hours, especially during the night time during which cell shutdown is triggered. As shown in Fig. \ref{fig:unavail_power}, during these hours the cell is unavailable and the transmit power drops. Likewise, the impact on downlink throughput and average number of users can be seen in Fig. \ref{fig:tp_users}. Next, we discuss how to analyze these KPIs together for the capacity and coverage cells to analyze the behaviour of energy saving features.

\subsection{Results}
In order to analyze the impact of energy saving using CCA, we take $p=3$ KPIs, including unavailable time, downlink maximum cell power, and average number of users for $c_x$ and construct data set $\mathbf{X}$. Likewise, $q=4$ KPIs selected for $c_y$ data set $\mathbf{Y}$ are downlink PRB utilization, uplink PRB utilization, throughput, and average number of users. After standardizing the variables in both sets, we run the CCA using Python's scikit-learn library to calculate the first pair of canonical variates. Correlation between first pair of canonical variates is $0.96925$, which shows that data sets are highly correlated. In order to understand the impact of constituent variables on canonical variate, it is important to compute the correlation between each canonical variate and its constituent variables. Correlations between variables of $\mathbf{X}$ and canonical variate of $\mathbf{X}$ are given in the Table \ref{tab:1}, where a:= unavailable time, b:= downlink maximum cell transmit power, and c:= average number of users. Unavailable time is negatively correlated as it decreases when other KPIs such as number of users or PRB usages increase.
\begin{table}[t]
\caption{Corr. between variables of $\mathbf{X}$ and 1st canon. variate}
\label{tab:1}
\centering
\begin{tabular}{|c||c||c||c||c|}
\hline
     & CCX 1 & a & b & c\\
\hline
CCX 1~& 1 & -0.30087 & 0.51371 & 0.99759   \\
\hline
a~& -0.30087 & 1 & -0.60886 & -0.32421     \\
\hline
b~& 0.51371   & -0.60886 & 1    & 0.50963     \\
\hline
c & 0.99759  & -0.32421 & 0.50963 & 1   \\
\hline
\end{tabular}
\end{table}
For data set $\mathbf{Y}$, we see in Table \ref{tab:2} that canonical variate has high correlation with all variables (a:=downlink PRB utilization, b:=uplink PRB utilization, c:=throughput, and d:=average number of users). Therefore, it is a good representation of coverage cell KPIs. In order to get further insight, we look at the correlations between each set of variables and the opposite group of canonical variates. Table \ref{tab:3} shows correlation between canonical variate of $\mathbf{X}$ and variables of $\mathbf{Y}$. Again, we can see high correlation with all the variables, especially for throughput and number of users. Finally, from Table \ref{tab:4} it is clear that average number of users is the capacity cell KPI that is strongly related to coverage cell KPIs. This makes sense because the number of users in capacity cell drops to zero when the cell is shutdown. Thus, the use of CCA enables us to understand how the two data sets are related and which KPIs in coverage cells are most impacted once the target capacity cell shutdown is triggered. 

\begin{table}[t]
\caption{Corr. between variables of Y and 1st canon. variate}
\label{tab:2}
\centering
\begin{tabular}{|c||c||c||c||c|c|}
\hline
     & CCY 1 & a & b & c & d\\
\hline
CCY 1~      & 1        & 0.58634  & 0.58688 & 0.72776  & 0.99481   \\
\hline
a~& 0.58634  & 1        & 0.35799 & 0.91020 & 0.65141     \\
\hline
b~& 0.58688  & 0.35799  & 1    & 0.43017   & 0.55117     \\
\hline
c~& 0.72776    & 0.91020  & 0.43017    & 1  & 0.77861    \\
\hline
d& 0.99481    & 0.651441 & 0.55117 & 0.77861 & 1   \\
\hline
\end{tabular}
\end{table}

\begin{table}[t]
\caption{Corr. between canon. variate of $\mathbf{X}$ and vars of $\mathbf{Y}$}
\label{tab:3}
\centering
\begin{tabular}{|c||c||c||c||c||c|}
\hline
     & CCX 1 & a & b & c & d\\
\hline
CCX 1~.      & 1        & 0.56831  & 0.56883 & 0.70538  & 0.96422   \\
\hline
a~& 0.56831  & 1        & 0.35799 & 0.91020 & 0.65141     \\
\hline
b~& 0.56883  & 0.35799  & 1    & 0.43017   & 0.55117     \\
\hline
c~& 0.70538    & 0.91020  & 0.43017    & 1  & 0.77861    \\
\hline
d& 0.96422    & 0.651441 & 0.55117 & 0.77861 & 1   \\
\hline
\end{tabular}
\end{table}

\begin{table}[t]
\caption{Corr. between canon. variate of $\mathbf{Y}$ and vars of $\mathbf{X}$}
\label{tab:4}
\centering
\begin{tabular}{|c||c||c||c||c|}
\hline
             & CCY 1    & a      & b      & c\\
\hline
CCY 1~      & 1        &-0.29164  & 0.49790  & 0.96692\\
\hline
a~        & -0.29164  & 1        & -0.60886 & -0.32421 \\
\hline
b~        & 0.49790  & -0.60886  & 1         & 0.50963     \\
\hline
c~        & 0.96692    & -0.32421  & 0.50963    & 1      \\
\hline
\end{tabular}
\end{table}

\section{Conclusions}
We have discussed the application of canonical correlation analysis (CCA) for radio access network data analysis, especially for performance data and key performance indicators (KPIs) in multivariate settings. CCA problem formulation for KPI analysis based on network data is discussed. Different data sets comprising of configuration and performance data are structured and mapped to a CCA problem, and subsequently solved for canonical correlation using available open source tools and libraries. As a case study, we discuss cell shutdown based energy saving use-case for LTE networks, where CCA is applied to understand the impact on capacity cell shutdown on coverage cell shutdown and relationships between different KPIs. Analysis of energy saving KPI data comprising of several counters related to downlink and uplink PRB usage, shutdown time, throughput, cell transmit power, and number of users, from a commercial LTE network show that CCA is an effective approach in analyzing and understanding the importance of different KPIs and their interrelationships at different levels. We conclude that due to its generic nature and the capability to formulate and solve a wide range of KPI analysis and performance monitoring problems comprising of multivariate data sets, the presented CCA based approach which is fundamentally well-suited for analyzing the performance impact and KPI behaviour in cases where multiple sources of related data need to be analyzed together for finding relationships and dependencies.




%







\end{document}